\documentclass[prl,twocolumn,superscriptaddress,showpacs,floatfix,longbibliography]{revtex4-1}
\usepackage{mathrsfs}
\usepackage{amssymb, amsbsy, amsmath, latexsym, dsfont, array, layout,
graphicx,mathrsfs,color,ulem,bm}

\begin{document}

\title{Direct Experimental Test of Forward and Reverse Uncertainty Relations}
\author{Lei Xiao}
\affiliation{Department of Physics, Southeast University, Nanjing 211189, China}
\affiliation{Beijing Computational Science Research Center, Beijing 100084, China}
\author{Bowen Fan}
\affiliation{Department of Physics, Southeast University, Nanjing 211189, China}
\author{Kunkun Wang}
\affiliation{Beijing Computational Science Research Center, Beijing 100084, China}
\author{Arun Kumar Pati}
\affiliation{Quantum Information and Computation Group, Harish-Chandra Research Institute, HBNI, Allahabad 211019, India}
\author{Peng Xue}\email{gnep.eux@gmail.com}
\affiliation{Beijing Computational Science Research Center, Beijing 100084, China}

\begin{abstract}
The canonical Robertson-Schr\"{o}dinger uncertainty relation provides a loose bound for the product of variances of two non-commuting observables. Recently, several tight forward and reverse uncertainty relations have been proved which go beyond the traditional uncertainty relations. Here we experimentally test multifold of state-dependent uncertainty relations for the product as well as the sum of variances of two incompatible observables for photonic qutrits. These uncertainty relations are independent of any optimization and still tighter than the Robertson-Schr\"{o}dinger uncertainty relation and other ones found in current literatures. For the first time, we also test the state-dependent reverse uncertainty relations for the sum of variances of two incompatible observables, which implies another unique feature of preparation uncertainty in quantum mechanics. Our experimental results not only foster insight into a fundamental limitation on preparation uncertainty but also may contribute to the study of upper limit of precession measurements for incompatible observables.
\end{abstract}

\maketitle

{\it Introduction:---} Uncertainty relations~\cite{H27,R29,WZ83,MP14,MBP17} are the hallmarks of quantum physics and have been widely investigated since their inception~\cite{ESS+12,SSE+13,SSD+15,PHC+11,RDM+12,WHP+13,RBB+14,KBOE14,WZB+16,XWZ+17,FWXX18,XFWX18}. These uncertainty relations impose fundamental limitation on the possible preparation of quantum states for which two non-commuting observables can have sharp values---often refereed as ``preparation'' uncertainty relations. They can be used in bounds for metrology~\cite{BC94,BCM96}, the security of quantum cryptography~\cite{FP96,RB09}, and the detection of nonclassical correlations~\cite{FT03,G04,CR07,ZXH+19}. Thus, the discovery of new uncertainty relations~\cite{MP14,MBP17} with tighter bounds has important potential implications for quantum information processing.

Uncertainty relations in terms of variances of incompatible observables are generally expressed in the product and the sum form. Both of these kinds of uncertainty relations express limitations in the possible preparations of the system by giving a lower bound to the product or sum of the variances of two observables. Most of the stronger uncertainty bounds~\cite{MP14,SQ16,YXWP15,XJJF16} depend on an orthogonal state to the state of the system. However, for higher dimensional systems, finding such an orthogonal state may be difficult. Recently, couple of state-dependent uncertainty relations with optimization free uncertainty bounds both in the sum and the product forms are derived by Mondal et al. in~\cite{MBP17}. These authors have also proved a state-dependent upper bound for the uncertainty relation which is tight. It is quite intriguing that the enshrined uncertainty relation due to  Robertson and Schr\"{o}dinger is a much weaker than the tight forward uncertainty relation proved in~\cite{MBP17}.

In this work, we report an experimental test of these new uncertainty relations with optimization free bounds and reverse uncertainty relations for single-photon measurements and demonstrate they are valid for states of a spin-$1$ particle. The experimental results we find agree well with the predictions of quantum theory and obey these new uncertainty relations. We realize a direct measurement model and give the first experimental investigation of the strengthened forward relations and the reverse uncertainty relation. Furthermore, in our experiment, every term  can be obtained directly by the outcomes of the projective measurements. Our test realizes a direct measurement model which leverages the requirement of quantum state tomography~\cite{XWZ+17,LXX+11,PHC+11}.

\begin{figure}
\centering
\includegraphics[width=0.5\textwidth]{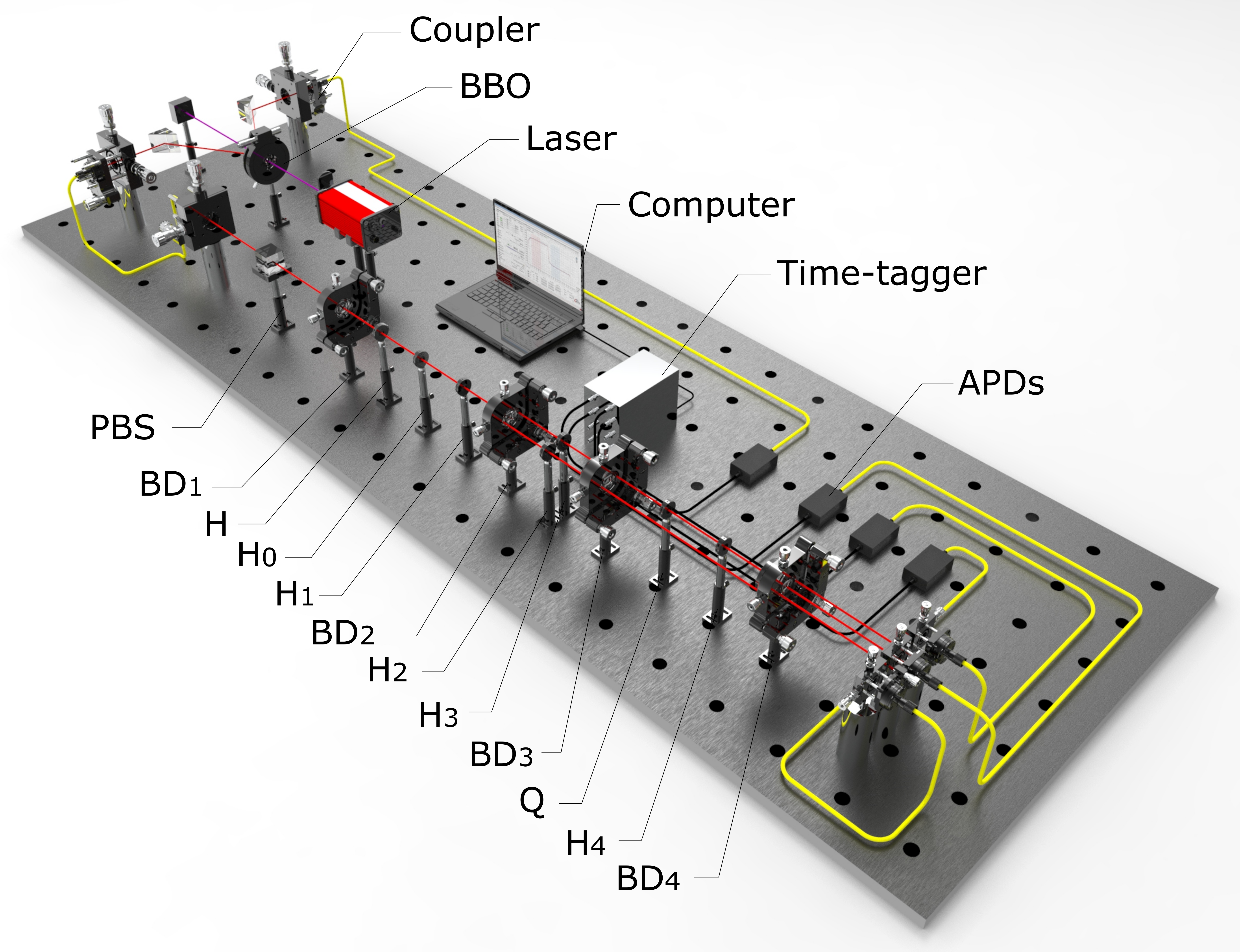}
\caption{Experimental setup for testing the uncertainty relations relating on both product and sum of variants of two incompatible observables $L_x$ and $L_y$ for the spin-$1$ particle with a state $|\Psi\rangle=(\cos\theta, -\sin\theta, 0)^\text{T}$ and reverse uncertainty relation. Photon pairs are generated via type-I SPDC. With the detection of the trigger, the heralded single photon is injected into the optical network. The PBS, HWPs (H and H$_0$) and BD$_1$ are used to generated the qutrit state $|\Psi\rangle$. The rest wave plates and BDs are used to realize the projective measurements of the observables $L_x$, $L_y$, $L_z$ and $L'$. The photons are finally detected by APDs in coincidence with the trigger photons.}
\label{fig1}
\end{figure}

\begin{figure}
\includegraphics[width=0.5\textwidth]{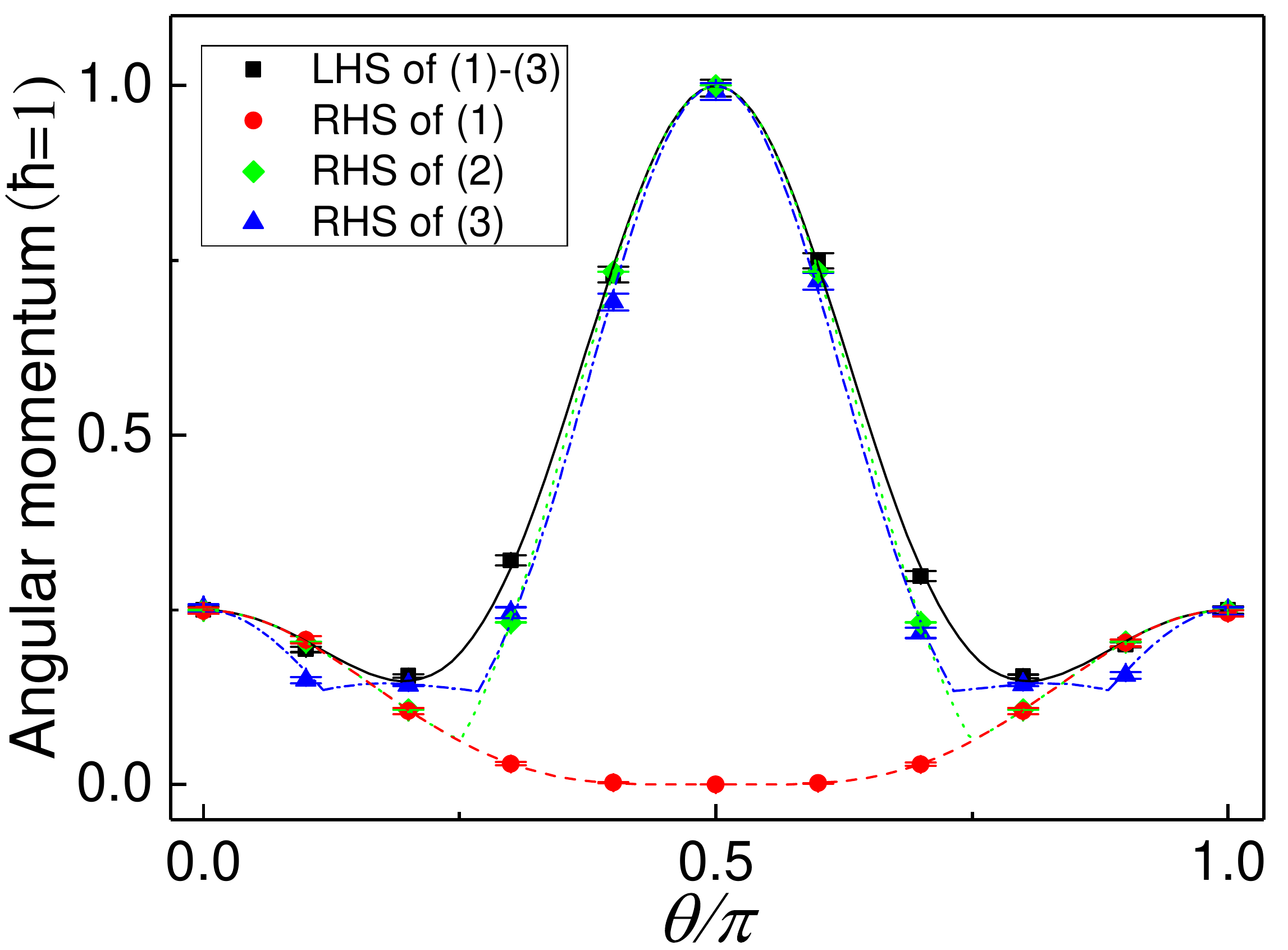}
   \caption{Experimental results of the uncertainty relations (\ref{eq:1})-(\ref{eq:3}) relating  the product of variances of two incompatible observables. Solid black curve and black squares represent theoretical predictions and experimental data of the left-hand side (LHS) of the inequalities (\ref{eq:1})-(\ref{eq:3}), i.e., $\Delta L^2_x\Delta L^2_y$ with $11$ states $|\Psi\rangle$. Red dots and dash curve represent the experimental results and theoretical predictions of the right-hand side (RHS) of inequality (\ref{eq:1}). Green diamonds and dotted curve represent the experimental results and theoretical predictions of the RHS of inequality (\ref{eq:2}). Blue triangles and dash-dotted curve represent the experimental results and theoretical predictions of the RHS of inequality (\ref{eq:3}). Error bars indicate the statical uncertainty which is obtained based on assuming Poissonian statistics.}
\label{fig2}
\end{figure}

\begin{figure}
\includegraphics[width=0.5\textwidth]{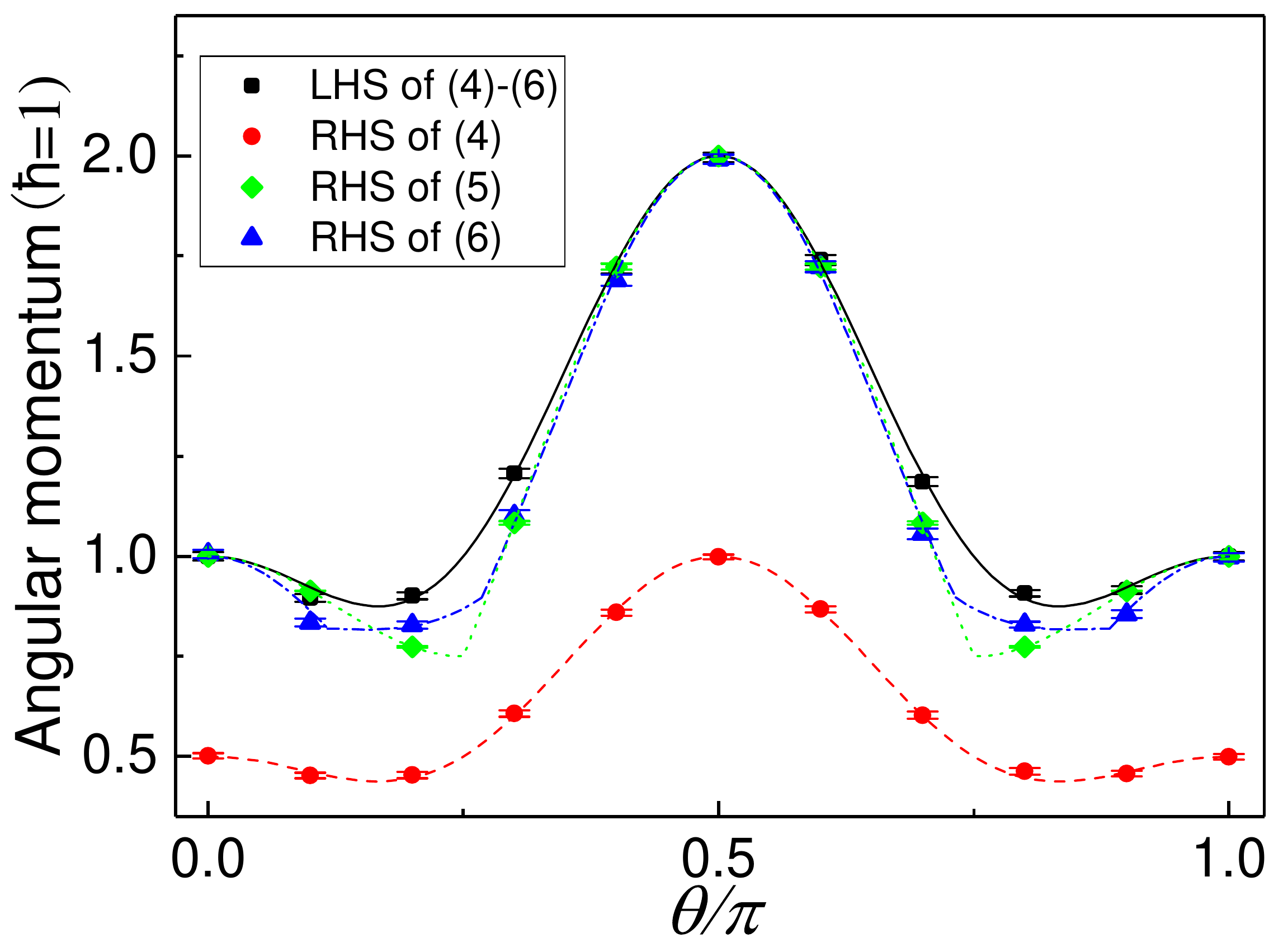}
   \caption{Experimental results of the uncertainty relations (\ref{eq:4})-(\ref{eq:5}) relating on the sum of variants of two incompatible observables. Solid black curve and black squares represent theoretical predictions and experimental data of the LHS of the inequalities (\ref{eq:4})-(\ref{eq:6}), i.e., $\Delta L^2_x+\Delta L^2_y$ with $11$ states $|\Psi\rangle$. Red dots and dashed curve represent the experimental results and theoretical predictions of the RHS of inequality (\ref{eq:4}). Green diamonds and dotted curve represent the experimental results and theoretical predictions of the RHS of inequality (\ref{eq:5}). Blue triangles and dash-dotted curve represent the experimental results and theoretical predictions of the RHS of inequality (\ref{eq:6}).}
\label{fig3}
\end{figure}

{\it Theoretical framework:---} Consider a quantum system that has been prepared in the state $|\Psi\rangle$ and we perform measurement of two incompatible observables $A$ and $B$.  The resulting bound on the product of uncertainties can be expressed as
\begin{eqnarray}
\hspace{-.2cm}\Delta A^2\Delta B^2\geq\left|\frac{1}{2}\langle[A,B]\rangle\right|
^2+\left|\frac{1}{2}\langle \{A,B\}\rangle-
\langle A\rangle\langle B\rangle\right|^2,
\label{eq:1}
\end{eqnarray}
where the expected values $\langle O\rangle = \langle \Psi|O| \Psi \rangle$ and variances $\Delta O^2=\langle O^2\rangle-\langle O\rangle^2$ are defined over an arbitrary state $|\Psi \rangle$ of the system. It is so-called Robertson-Shr\"{o}dinger uncertainty relation~\cite{E36}. This uncertainty relation is well known, however, its bound is not optimal. In~\cite{MBP17}, an alternative uncertainty relation with a tighter bound is provided
\begin{eqnarray}
\Delta A^{2}\Delta B^{2} \geqslant \Big(\sum_{n}|\langle\Psi|\overline{A}|\psi_{n}\rangle\langle\psi_{n}|\overline{B}|\Psi\rangle|\Big)^2,
\label{eq:2}
\end{eqnarray}
where $A=\sum_{i}a_{i}|a_{i}\rangle\langle a_{i}|$ and $B=\sum_{i}b_{i}|b_{i}\rangle\langle b_{i}|$, $\overline{A}=(A-\langle A\rangle)=\sum_{i}\tilde{a}_i|a_i\rangle\langle a_i|$
and $\overline{B}=(B-\langle B\rangle)=\sum_{i}\tilde{b}_i|b_i\rangle\langle b_i|$, $\{|\psi_{n}\rangle\}$ is an arbitrary complete basis. Though the bound of the new uncertainty relation is indeed tighter than that of the Robertson-Shr\"{o}dinger uncertainty relation, an optimization over different bases is required for the tightest bound.

Furthermore, an optimization-free uncertainty relation for two incompatible observables is derived in~\cite{MBP17} which is given by
\begin{eqnarray}
&\Delta A^2\Delta B^2\geq
\Bigg(\sum_{i}\sqrt{F_{\Psi}^{a_{i}}}\sqrt{F_{\Psi}^{b_{i}}}
\tilde{a}_{i}\tilde{b}_{i}\Bigg)^2,
\label{eq:3}
\end{eqnarray}
where $F_{\Psi}^{x}=|\langle\Psi|x\rangle|^2$ is the fidelity between $|\Psi \rangle$ and $|x \rangle=|a_i\rangle,|b_i\rangle$. This uncertainty relation depends on the transition probability between the state of the system $|\Psi\rangle$ and the eigenstates of the observables $|a_i\rangle$ and $|b_i\rangle$. The incompatibility is captured here not by the noncommutativity, but rather by the nonorthogonality of the state of the system and the eigenstates of the observables. The bound of this uncertainty relation~(\ref{eq:3}) is tighter than the other bounds most of the time and even surpasses the bound given by (\ref{eq:2}) without any optimization.

To fully capture the concept of incompatible observables, an uncertainty relation proposed relating on that sum of variances of two incompatible observables is derived in~\cite{MP14}
\begin{eqnarray}
\Delta A^2+\Delta B^2
&\geq &\frac{1}{2}\Delta (A+B)^2.
\label{eq:4}
\end{eqnarray}
An optimization over a set of states $|\psi_n\rangle$ provides a more tighter bound as
\begin{eqnarray}
\Delta A^2+\Delta B^2&\geq & \frac{1}{2}\sum_{n}\Big(|\langle\psi_{n}|\overline{A}|\Psi\rangle|+|\langle\psi_{n}|\overline{B}|\Psi\rangle|\Big)^2.
\label{eq:5}
\end{eqnarray}
Furthermore, an uncertainty relation for the sum of variances of two incompatible observables with optimization-free bound is also derived in~\cite{MBP17}
\begin{eqnarray}
\Delta A^2+\Delta B^2
&\geq &\frac{1}{2}\sum_{i}\Big(\tilde{a}_{i}\sqrt{F_{\Psi}^{a_{i}}}+
\tilde{b}_i\sqrt{F_{\Psi}^{b_{i}}}\Big)^2.
\label{eq:6}
\end{eqnarray}

The standard preparation uncertainty relations---forward URs provide lower bound to the product or the sum of variances. However, quantum theory also restricts how large the variances can be. The upper bound to the sum of variances of two incompatible observables is expressed by the reverse uncertainty relation~\cite{MBP17}
\begin{equation}
\Delta A^2+ \Delta B^2\leq \frac{2\Delta (A-B)^2}{\Big [1-\frac{\text{Cov}(A,B)}{\Delta A\Delta B}\Big ]}-2\Delta A\Delta B,
\label{eq:7}
\end{equation}
where $\text{Cov}(A,B)=\frac{1}{2}\langle\{ A,B\}\rangle-\langle A\rangle\langle B\rangle$ is the quantum covariance of the operators $A$ and $B$. Thus, for two incompatible observables, the forward and the reverse uncertainty relations set fundamental zone within which the quantum fluctuations are restricted, i.e., the intrinsic uncertainties cannot be too small and too large.

\begin{figure}
\includegraphics[width=0.5\textwidth]{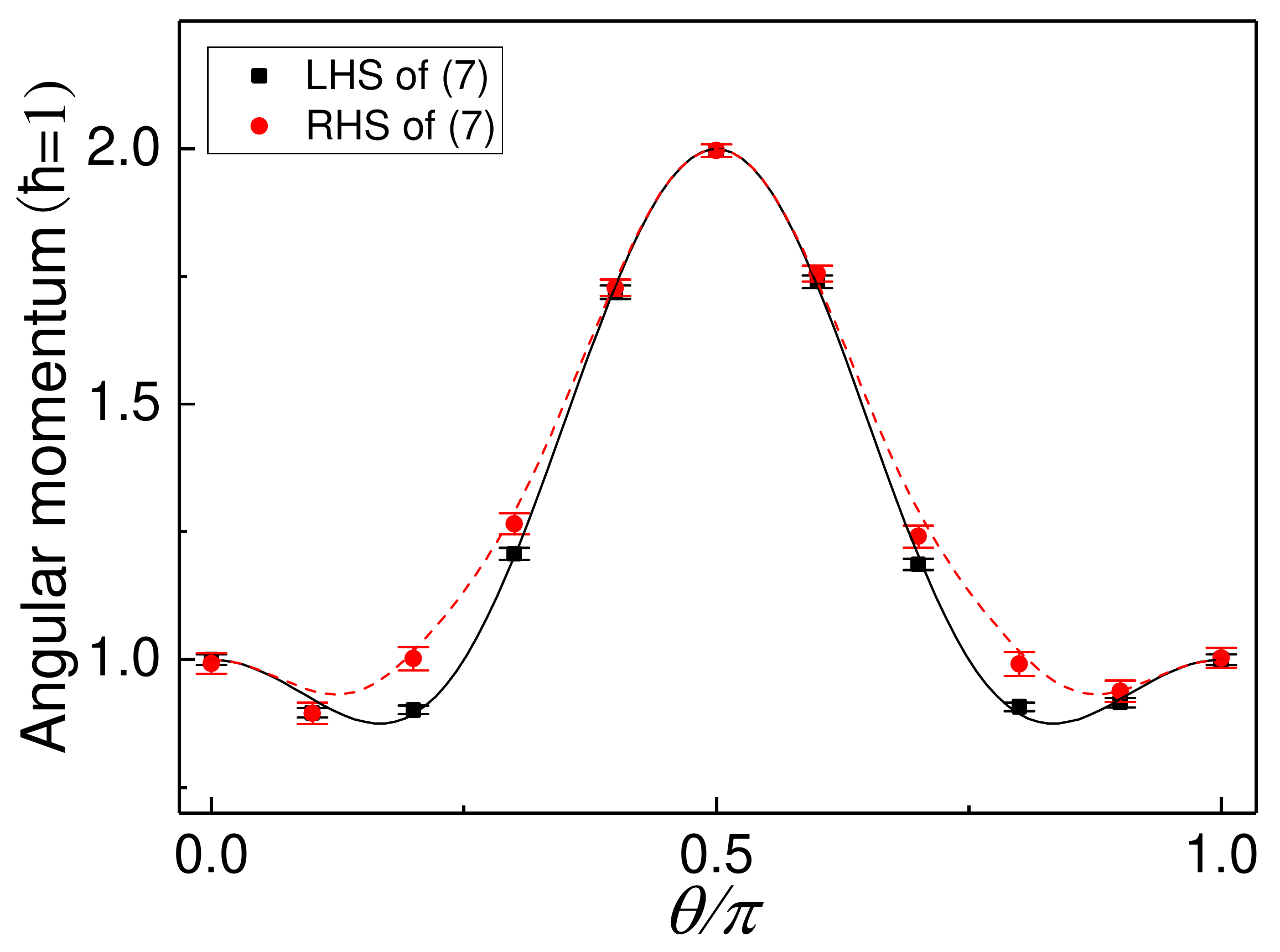}
   \caption{Experimental results of the reverse uncertainty relation. Theoretical predictions of LHS and RHS of the inequality (\ref{eq:7}) are represented by the curves. Whereas experimental data are indicated by the symbols.
   }
\label{fig4}
\end{figure}

{\it Experimental investigation:---}To Test the uncertainty relations (\ref{eq:1})-(\ref{eq:6}) and the reverse uncertainty relation (\ref{eq:7}),
we choose two components of the angular momentum for spin-$1$ particle as two observables:
\begin{equation}
L_x= \frac{1}{\sqrt{2}}\begin{pmatrix}
0 & 1 & 0 \\
1 & 0 & 1 \\
0 & 1 & 0 \\
\end{pmatrix},
L_y= \frac{1}{\sqrt{2}}\begin{pmatrix}
0 & -i & 0 \\
i & 0 & -i \\
0 & i & 0 \\
\end{pmatrix}.
\end{equation}
For convenience, we also define an observable as $L'=L_xL_y+L_yL_x=\begin{pmatrix}
0 & 0 & -i \\
0 & 0 & 0 \\
i & 0 & 0 \\
\end{pmatrix}$ and the other component of the angular momentum $L_z=-i\left[L_x,L_y\right]=\begin{pmatrix}
1 & 0 & 0 \\
0 & 0 & 0 \\
0 & 0 & -1 \\
\end{pmatrix}$ is used. Then the inequalities can be rewritten and both left- and right-hand sides can be measured directly.

The inequality (\ref{eq:1}) can be rewritten as
\begin{equation*}
\Delta L_x^2\Delta L_y^2\geq\left|\frac{1}{2}\langle L_z\rangle\right|
^2+\left|\frac{1}{2}\langle L' \rangle-
\langle L_x\rangle\langle L_y\rangle\right|^2.
\end{equation*}
All the lift- and right-hand sides of the inequalities are expected values of the obserables $L_i$ ($i=x,y,z$) and $L'$ and can be read out from the experimental results.

The inequality (\ref{eq:2}) can be rewritten as
\begin{equation*}
\Delta L_x^2\Delta L_y^2\geq\left(\sum_{n} |C_n-D_n \langle L_x\rangle| \right)^2,
\end{equation*}
where $C_n=\text{Tr}\left(\rho L_x |\psi_n\rangle \langle \psi_n|L_y\right)$, $D_n=\text{Tr}\left(\rho |\psi_n\rangle \langle \psi_n|L_y\right)$,
$\rho=|\Psi\rangle\langle\Psi|$, and $|\psi_1\rangle=(1,0,0)^\text{T}$, $|\psi_2\rangle=(0,1,0)^\text{T}$ and $|\psi_3\rangle=(0,0,1)^\text{T}$ are the
mutual orthogonal basis vectors for ${\cal H}^3$.

The inequality (\ref{eq:3}) can be rewritten as
\begin{equation*}
\Delta L_x^2\Delta L_y^2
\geq \Big(\sum_{i}\tilde{a}_{i} \tilde{b}_i \sqrt{F_{\Psi}^{a_{i}}}
\sqrt{F_{\Psi}^{b_{i}}}\Big)^2,
\end{equation*}
where $F_{\Psi}^{a_i(b_i)}=\left|\langle \Psi|a_i(b_i)\rangle\right|^2$, $|a_i(b_i)\rangle$ indicates the eigenstate of $L_x(L_y)$ with the
eigenvalue $a_i(b_i)=-1,1,0$, and $\tilde{a}_{i}(\tilde{b}_{i})=a_i(b_i)-\langle L_x(L_y) \rangle$.

The inequality (\ref{eq:4}) can be rewritten as
\begin{align*}
\Delta L_x^2+\Delta L_y^2\geq \langle L_x^2 \rangle + \langle L_y^2 \rangle+\langle L' \rangle-\Big(\langle L_x \rangle+\langle L_y \rangle\Big)^2.
\end{align*}

The inequality (\ref{eq:5}) can be rewritten as
\begin{equation*}
\Delta L_x^2+\Delta L_y^2\geq \sum_{n} \left(|E_n-F_n \langle L_x\rangle|+|G_n| \right)^2,
\end{equation*}
where the coefficients are $E_n=\langle \psi_n |L_x|\Psi \rangle$, $F_n= \langle \psi_n |\Psi\rangle$ and $G_n=\langle \psi_n |L_y|\Psi \rangle$.

The inequality (\ref{eq:6}) can be rewritten as
\begin{equation*}
\Delta L_x^2+\Delta L_y^2
\geq \frac{1}{2}\sum_{i}\Big(\tilde{a}_{i}\sqrt{F_{\Psi}^{a_{i}}}+
\tilde{b}_i\sqrt{F_{\Psi}^{b_{i}}}\Big)^2,
\end{equation*}
where the coefficients $F_\Psi^{a_i(b_i)}$, $\tilde{a}_i(\tilde{b}_i)$ and $a_i(b_i)$ are defined in rewritten inequality (\ref{eq:3}).

The reverse uncertainty relation (\ref{eq:7}) can be rewritten as
\begin{equation*}
\Delta L_x^2+ \Delta L_y^2\leq \frac{2\Delta (L_x-L_y)^2}{\Big [1-\frac{\langle L' \rangle/2- \langle L_x \rangle \langle L_y \rangle}{\Delta L_x\Delta L_y}\Big ]}-2\Delta L_x \Delta L_y,
\end{equation*}
where $\Delta (L_x-L_y)=\langle L_x^2 \rangle + \langle L_y^2 \rangle - \langle L' \rangle-(\langle L_x \rangle-\langle L_y \rangle)^2$.

Thus, all terms in both lift- and right-hand sides of the uncertainty relations and reverse one are related to expected values of the obserables $L_i$ ($i=x,y,z$) and $L'$ and can be read out directly from the outcomes of the projective measurements.

{\it Experimental investigation:---}We report the experimental test of these uncertainty relations (\ref{eq:1})-(\ref{eq:6}) and the reverse uncertainty relation (\ref{eq:7}) for a single-photon measurement. The experimental setup shown in Fig.~\ref{fig1} involves two stages: state preparation and projective measurement.

In our experiment, a photonic qutrit is encoded in three modes and the basis states are
$\{|\psi_1\rangle,| \psi_2 \rangle, | \psi_3 \rangle\}=\{(1,0,0)^\text{T},(0,1,0)^\text{T},(0,0,1)^\text{T}\}$,
which indicate the states of the horizontally polarized photons in the upper spatial mode, the vertically polarized photons in the upper spatial mode, and the vertically polarized photons in the lower spatial mode, respectively. Polarization-degenerated photon pairs are produced in a type-I spontaneous parametric down-conversion (SPDC) with the help of an interference filter which restricts the photon bandwidth to $3$nm. This trigger-herald pair is registered by a coincidence count at two single-photon avalanche photodiodes (APDs) with $3$ns time window. Total coincidence counts are about $10^4$.

In the state preparation stage, the heralded single photons pass through a polarizing beam splitter (PBS) and are split into two parallel spatial modes---upper and lower modes by a beam displacer (BD) whose optical axis is cut so that vertically polarized photons are directly transmitted and horizontal photons undergo a lateral displacement into a neighboring mode. Then, two half-wave plates (HWPs) H with a certain setting angle $\theta/2$ and H$_0$ at $0$ are applied on the photons in the upper mode. The matrix form of the operation of HWP with the setting angle $\vartheta$ is $\begin{pmatrix}
    \cos 2\vartheta & \sin 2\vartheta \\
    \sin 2\vartheta & -\cos 2\vartheta \\
    \end{pmatrix}$. We prepare a family of qutrit states $|\Psi\rangle=(\cos\theta, -\sin\theta, 0)^\text{T}$ as the system state, where $\theta=j\pi/10$ ($j=0,\cdots,10$). Thus totally $11$ states are chosen for testing the uncertainty relations (\ref{eq:1})-(\ref{eq:6}) and the reverse one (\ref{eq:7}).

To test the uncertainty relations (\ref{eq:1})-(\ref{eq:6}) and the reverse one (\ref{eq:7}) which can be rewritten and only depend on the expected values of the observables $L_i$ ($i=x,y,z$) and $L'$. An arbitrary observable $M$ can be written as $M=\sum_{i} m_i |m_i\rangle \langle m_i|$, where $|m_i\rangle$ is an eigenstate of $M$ and $m_i$ is the corresponding eigenvalue. The expected value of $M$ is
\begin{align}
\langle M \rangle &= \langle \Psi| M | \Psi\rangle=\sum_{i} m_i \langle \Psi| m_i \rangle \langle m_i | \Psi \rangle\\ \nonumber
&=\sum_{i} m_i \left| \langle \Psi| m_i \rangle
\right|^2.
\end{align}

A unitary operator is defined as $U=\sum_i|i\rangle \langle m_i|$ and is applied on the system in the initial state $|\Psi\rangle$ which is then projected into the basis state $|i\rangle$ ($i=0,1,2$). The overlap $|\langle \Psi|m_i\rangle|^2=\text{Tr}\left(|\Psi\rangle\langle \Psi| U^\dagger |i\rangle\langle i| U\right)$ equals to the probability $P_i$ of the photons being measured in the basis state $|i\rangle$.

For example, corresponding to the observable $L_x$, the unitary operator is defined
\begin{equation}
U=\begin{pmatrix}
\frac{1}{2} & -\frac{1}{\sqrt{2}} & \frac{1}{2} \\
\frac{1}{2} & \frac{1}{\sqrt{2}} & \frac{1}{2} \\
-\frac{1}{\sqrt{2}} & 0 & \frac{1}{\sqrt{2}} \\
\end{pmatrix},
\end{equation}
which can be further decomposed into
\begin{align}
U&=U_3 U_2 U_1\\ & =\begin{pmatrix}
 \frac{1}{\sqrt{2}} & \frac{1}{\sqrt{2}} & 0 \\
-\frac{1}{\sqrt{2}} & \frac{1}{\sqrt{2}} & 0 \\
0 & 0 & 1\\
\end{pmatrix}
\begin{pmatrix}
-1 & 0 & 0 \\
0 & \frac{1}{\sqrt{2}} & \frac{1}{\sqrt{2}} \\
0 & \frac{1}{\sqrt{2}} & \frac{1}{\sqrt{2}} \\
\end{pmatrix}
\begin{pmatrix}
0 & 1 & 0 \\
1 & 0 & 0 \\
0 & 0 & 1 \\
\end{pmatrix}.\nonumber
\end{align}

Thus in the measurement stage, the above three unitary operators $U_{1,2,3}$ can be realized via the optical circuit in Fig.~\ref{fig1}. Each of them applies a rotation on just two of the basis states, leaving the other unchanged. For example, $U_1$ is realized by a HWP (H$_1$ at $45^\circ$) applying on the photons in the upper mode, which exchanges the polarizations of the photons in the upper mode and keeps the state of the photons in the lower mode unchanged. Similarly, $U_3$ is realized via a quarter-wave plate (QWP, Q) and a HWP (H$_4$) applying on the photons in the upper mode, which implement a rotation on the polarization states of the photons in the upper mode and keeps the state of the photons in the lower mode unchanged. Whereas, $U_2$ is realized by two BDs and two HWPs (H$_2$ and H$_3$). The BDs are used for basis transformation and the HWPs are used to apply a rotation on the polarization states. The last BD is used to project the photons into the basis states $|i\rangle$ ($i=0,1,2$). The probability of the photons being measured in $|i\rangle$ is obtained by normalizing photon counts in the certain spatial mode to total photon counts. Angles of the wave plates are shown in Table~\ref{table1}.

\begin{table}[h]
\caption{The setting angles of the wave plates for the projective measurement of the observables $L_x$, $L_y$, $L_z$ and $L'$.  Here ``$-$'' denotes the corresponding wave plate is removed from the optical circuit.}
\begin{tabular}{c||c|c|c|c|c}
\hline
Observations & H$_{1}$ & H$_{2}$ & H$_{3}$ & H$_{4}$ & Q\\
\hline
$L_x$ & $45.00^{\circ}$ & $22.50^{\circ}$  & $-45.00^{\circ}$ & $22.50^{\circ}$ & $-$\\
\hline
$L_y$  & $-45.00^{\circ}$ & $22.50^{\circ}$ & $45.00^{\circ}$ & $22.50^{\circ}$ & $0$\\
\hline
$L_z$  & $-$ & $-$ & $-$ & $-$ & $-$\\
\hline
$L'$ &  $0$ & $0$  & $-45.00^{\circ}$ & $22.50^{\circ}$ & $0$\\
\hline
\end{tabular}
\label{table1}
\end{table}

{\it Experimental results:---}In Fig.~\ref{fig2}, we show the direct demonstration of the uncertainty relations (\ref{eq:1})-(\ref{eq:3}) related to product of variances of two incompatible observables for photonic qutrits. The bound given by (\ref{eq:2}) is one of the tightest bounds but to achieve it optimization is required. Whereas, the bound given by (\ref{eq:3}) is tighter than the other bounds most of the time and even surpasses the optimized bound, yet it does not need any optimization. Both bounds in (\ref{eq:2}) and (\ref{eq:3}) are tighter than that given by the Schr\"{o}dinger uncertainty relation (\ref{eq:1}).

As shown in Fig.~\ref{fig3}, the bound obtained in (\ref{eq:6}) is one of the tightest optimization-free bounds. Whereas, the bound given by (\ref{eq:5}) requires optimization over the states perpendicular to the chosen state of the system and is surpassed for only a few states of the system. Both bounds in (\ref{eq:5}) and (\ref{eq:6}) are tighter than that given by (\ref{eq:4}). Our experimental data fit the theoretical predictions and satisfy the uncertainty relations of either product of sum of variances of two incompatible observables.

In Fig.~\ref{fig4}, we show the experimental demonstration of the reverse uncertainty relation (\ref{eq:7}). For some states, the inequality (\ref{eq:7}) becomes an equality, which means the reverse uncertainty relation is tight. For the coefficients of the system state $\theta=0,\pi/2,\pi$, the experimental results of the LHS and RHS of (\ref{eq:7}) are $\{0.99275\pm0.01985,0.99981\pm 0.01038\}$, $\{1.99614\pm0.01197,1.99622\pm0.01214\}$ and $\{0.99988\pm0.01028,1.00343\pm0.01967\}$, respectively, which agree with their theoretical predictions $1,2,1$ very well.

{\it Conclusion:---}The uncertainty relations are the most fundamental relations in quantum theory. A correct understanding and experimental confirmation of uncertainty relations will not only foster insight into foundational problems but also advance the precision measurement technology in quantum information processing. In this work, we have experimentally tested several forward as well as reverse state-dependent uncertainty relations for the product as well as the sum of variances of two incompatible observables for photonic qutrits. These uncertainty relations are independent of any optimization and still tighter than the Robertson-Schr\"{o}dinger uncertainty relation and other ones existing in the current literatures. We have also tested, for the first time, the state-dependent reverse uncertainty relations for the sum of variances of two incompatible observables, which implies an another unique feature of quantum mechanics. The experimental test of the forward and the reverse uncertainty relations vividly demonstrates that quantum fluctuations do remain within the quantum tract due to the incompatible nature of the observables.

The fruition of our experiment relies on a stable interferometric network with simple linear optical elements. Though both sides of these inequalities can be calculated from the density matrices of the system states which are reconstructed by quantum state tomography. In our experiment, every term of these inequalities can be obtained directly by the outcomes of the projective measurements, and the experimental results are in a perfect agreement with theoretical predictions. Our test realizes a direct measurement model which much simplifies the experimental realization and leverages the requirement of quantum state tomography. Our experimental results not only provide deep insights into fundamental limitations of measurements but also may contribute to the study an upper time limit of quantum evolutions in future.

\begin{acknowledgments}
This work has been supported by the National Natural Science Foundation of China (Grant Nos. 11674056 and U1930402), the startup funding of Beijing Computational Science Research Center, and Postgraduate Research \& Practice Innovation Program of Jiangsu Province (Grant No. KYCX18\_0056).
\end{acknowledgments}

\bibliography{referencebib}

\end{document}